# Bifrost: Reinventing WiFi Signals Based on Dispersion Effect for Accurate Indoor Localization

Yimiao Sun, Yuan He*, Jiacheng Zhang, Xin Na, Yande Chen, Weiguo Wang, Xiuzhen Guo
School of Software and BNRist, Tsinghua University
sym21@mails.tsinghua.edu.cn, heyuan@tsinghua.edu.cn
{zhangjc21,nx20,cyd22,wwg18}@mails.tsinghua.edu.cn, guoxiuzhen94@gmail.com

## ABSTRACT

WiFi-based device localization is a key enabling technology for smart applications, which has attracted numerous research studies in the past decade. Most of the existing approaches rely on Line-of-Sight (LoS) signals to work, while a critical problem is often neglected: In the real-world indoor environments, WiFi signals are everywhere, but very few of them are usable for accurate localization. As a result, the localization accuracy in practice is far from being satisfactory. This paper presents Bifrost, a novel hardware-software co-design for accurate indoor localization. The core idea of Bifrost is to reinvent WiFi signals, so as to provide sufficient LoS signals for localization. This is realized by exploiting the dispersion effect of signals emitted by the leaky wave antenna (LWA). We present a low-cost plug-in design of LWA that can generate orthogonal polarized signals: On one hand, LWA disperses signals of different frequencies to different angles, thus providing Angle-of-Arrival (AoA) information for the localized target. On the other hand, the target further leverages the antenna polarization mismatch to distinguish AoAs from different LWAs. In the software layer, fine-grained information in Channel State Information (CSI) is exploited to cope with multipath and noise. We implement Bifrost and evaluate its performance under various settings. The results show that the median localization error of Bifrost is 0.81m, which is 52.35% less than that of SpotFi, a state-of-the-art approach. SpotFi, when combined with Bifrost to work in the realistic settings, can reduce the localization error by 33.54%.

## CCS CONCEPTS

· **Networks** → Location based services; · **Information systems** → Location based services;

## KEYWORDS

WiFi Localization, Indoor Localization, Leaky Wave Antenna, RF Computing

---

‡Yuan He is the corresponding author.

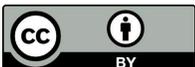

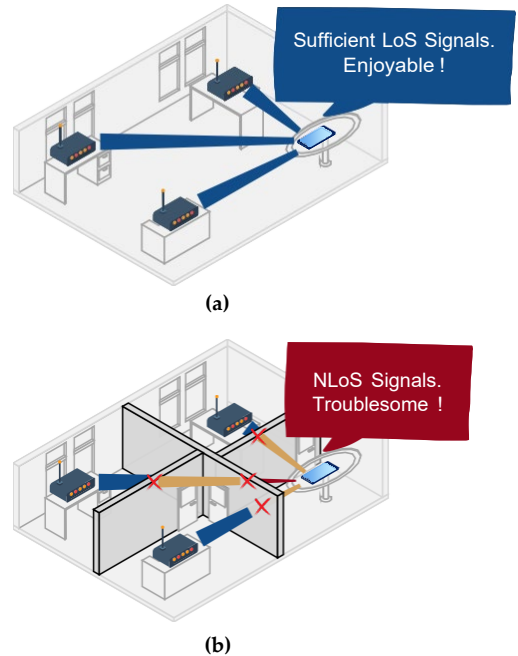

**Figure 1: A model-driven method works well when (a) sufficient LoS signals are available but becomes inaccurate when (b) NLoS signals have to be used.**

## 1 INTRODUCTION

Location information [32, 63, 79, 80] is crucial, especially for smart indoor applications [50, 60, 67, 72], such as smart home [54, 62], indoor navigation [7, 18, 19, 59] and so on. Due to the ubiquitous deployment of WiFi access points (APs) and wide availability of WiFi modules on the devices, WiFi-based localization [16, 25, 49, 56, 57, 61, 64, 65, 68–70, 73, 74, 82] appears to be promising for indoor localization. The existing works of WiFi-based indoor localization can be broadly grouped into two categories, data-driven methods and model-driven methods.

Data-driven methods are typically represented by fingerprint [14, 44, 61]. These methods need to collect Received Signal Strength (RSS) or CSI at different places to construct a database mapping RSS (or CSI) with locations, which is a labor-intensive process. Also, their performance may be vulnerable to dynamic environments.

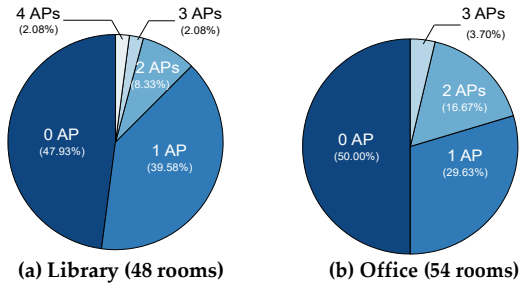

**Figure 2: The number of LoS APs in each room in a library and an office building.**

Model-driven methods induce less labor cost and attract more research studies. Generally, a model-driven method calculates the location by estimating signals' Angle-of-Arrival (AoA) [2, 23, 24, 69], Time-of-Flight (ToF) [70, 81] or both [9, 16, 43]. Most of the existing approaches rely on Line-of-Sight (LoS) signals to work, as Fig. 1(a) illustrates, while a critical problem is often neglected: In the real-world indoor environments, WiFi signals are everywhere, but very few of them are usable for accurate localization. As an example to validate this finding, Fig. 2 plots the statistics of the real deployment of WiFi APs in a library (48 rooms) and an office building (54 rooms). The data shows that in nearly half of all the rooms, there is not even one LoS AP available. The rooms with sufficient LoS signals account for less than 5% of all the rooms. In other words, the chance for a WiFi device to receive sufficient LoS WiFi AP signals, namely the case for it to be accurately localized by using an existing approach, is less than 5%. That well explains why the practical performance of using the existing localization approaches is far from being satisfactory.

A straightforward idea to address the above problem is to increase the number of deployed WiFi APs, until everywhere is covered by at least 3 LoS APs. It isn't practical, however. Taking the library and office building investigated in Fig. 2 as an example, typically there are 50 rooms in a building. Covering every room with 3 APs requires 150 APs to be deployed, which means multiple drawbacks, such as substantial deployment cost of cables (connecting the APs), overly crowded wireless spectrum, and frequent interference and collisions in the wireless communication.

This paper presents a novel approach called Bifrost[1], a plug-and-play and cost-effective scheme to significantly enhance the availability of LoS WiFi signals and in turn the localization accuracy. In light of the research progress on leaky wave antenna (LWA) in recent years [21, 22, 42, 47, 48, 76], Bifrost exploits dispersion effect of wireless signals [33]. Deployed in the space covered by WiFi signals, a LWA can receive those signals and then radiate them at different frequencies towards different directions, exhibiting frequency and spatial division multiplexing (FSDM) features, as is reinventing[2] WiFi signals.

---
[1] In Norse mythology, Bifrost is a rainbow bridge that reaches between Midgard (Earth) and Asgard (the realm of gods).
[2] The word "reinventing" means that Bifrost makes WiFi signals look different from their original form by using the LWAs. The signal emitted by the LWAs has two new properties, dispersion effect and circular polarization.

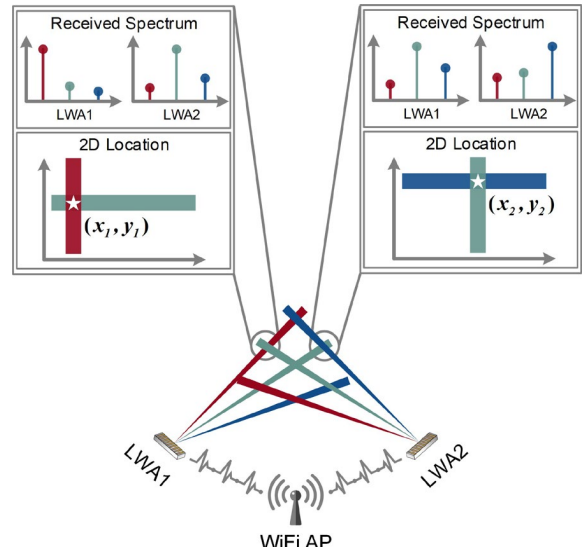

**Figure 3: The high-level principle of Bifrost.**

Fig. 3 illustrates the high-level principle of Bifrost. To localize a target device, Bifrost uses two LWAs to transform WiFi signals into FSDM signals, so the target device will receive two LoS FSDM signals with a unique pair of frequencies. Since the frequency and the propagation direction of FSDM signals are coupled, the target device can estimate its AoAs to both LWAs by analyzing the received spectrum and then calculate its location.

Compared with using WiFi APs, using LWA to assist localization offers the following two distinct advantages:

1) **Cost-effective.** The cost of a LWA in Bifrost is 7.41 USD (4.36 USD for the material cost and 3.05 USD for the control module), which is significantly lower than that of a WiFi AP (typically 30 ∼ 100 USD [3–6]).

2) **Easy to Use.** Deploying a LWA is very convenient. It can operate in a plug-and-play manner without the need for connecting power cables.

Leveraging these two advantages, Bifrost can be easily implemented in any environment with WiFi coverage, no matter whether the WiFi signals are LoS or not. Bifrost can either work independently, or cooperatively with other conventional WiFi-based localization methods.

The design of Bifrost tackles several critical challenges, which are summarized as follows:

**Ambiguity between Different LWAs.** As Fig. 3 shows, a target device may receive signals from two LWAs, which are reinvented from the same WiFi signal source. Without a special design, it is almost impossible for the target to distinguish one LWA from the other. To overcome this problem, the LWAs in Bifrost are designed to generate orthogonal circular polarized (CP) signals, so that they won't mix up with each other (§3.1). Polarization of LWA signals can be conveniently switched by altering the input port of WiFi signals, without the need for reconstruction or modifications to the LWA's structure.

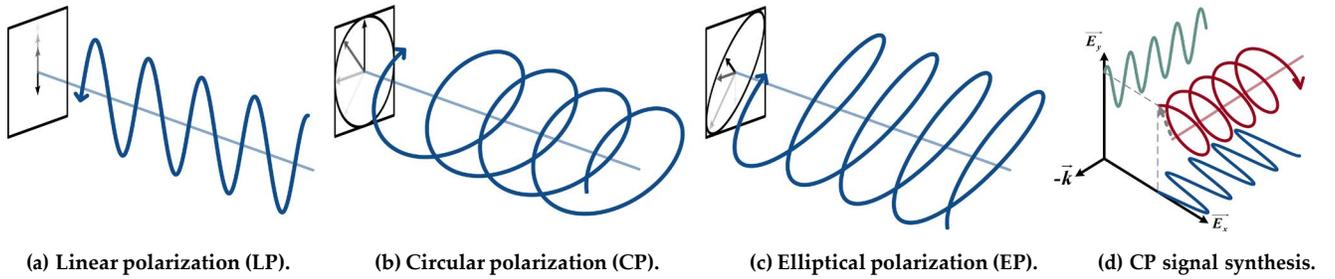

**(a) Linear polarization (LP).**  **(b) Circular polarization (CP).**  **(c) Elliptical polarization (EP).**  **(d) CP signal synthesis.**

**Figure 4: The properties of polarized electromagnetic waves.**

**Signal Extraction from the Interfered Frequency Band.** Since FSDM signals radiated by LWAs are transformed from existing WiFi signals, the two types of signals operate within the same frequency band and can be simultaneously received by a target device. Directly using such signals leads to erroneous AoA estimation. To deal with such interference, LWAs in Bifrost work in a duty-cycled manner. The target device is able to detect distinctive variation of the signal amplitude at the frequencies of FSDM signals (§3.3). By analyzing WiFi CSI, the target device can effectively extract the desired FSDM signals from the interfered frequency band.

**Indoor Multipath Effect.** The multipath effect in the indoor environment may seriously affect the quality of the received FSDM signals and further affect the localization accuracy. In order to identify FSDM signals propagating along the LoS path, Bifrost operates in two steps. First, we map frequencies of FSDM signals with subcarriers in CSI and cluster adjacent subcarriers to only retain the cluster with the highest energy (§3.4). Second, we take the intersection of two clusters (corresponding to the two orthogonal CP signals), and determine the final frequency by weighting the center frequency of the remaining clustered subcarriers (§3.5).

Our contributions can be summarized as follows:

1) We tackle a significant problem, namely the limited availability of LoS signals, which is overlooked by the existing works on WiFi-based indoor localization. We reinvent WiFi signals by exploiting the dispersion effect, which represents a new direction of utilizing LWAs.

2) We address a series of non-trivial challenges, such as signal ambiguity, interference, and multipath effect, *etc.* The design of Bifrost effectively ensures the quality of signals used for localization.

3) We implement Bifrost and evaluate its performance under various settings. The results show that the median localization error of Bifrost is 0.81m, which is 52.35% less than that of SpotFi, a state-of-the-art approach. SpotFi, when combined with Bifrost to work in the realistic settings, can reduce the localization error by 33.54%.

This paper proceeds as follows: §2 introduces background knowledge on the signal polarization and the LWA. Then §3 unfolds the design of Bifrost in both hardware and software. The implementation and evaluation results are presented in §4. We discuss practical issues in §5 and summarize related works in §6. This work is concluded in §7.

## 2 PRIMER

This section introduces preliminary knowledge of our work: polarization of wireless signals and leaky wave antenna.

### 2.1 Signal Polarization

Polarization is a fundamental property of wireless signals, including FSDM and WiFi signals investigated in this work. It represents the direction of the signal's electric field, which can be denoted as $\vec{E}$ and can be decomposed into the horizontal component $\vec{E_x}$ and the vertical component $\vec{E_y}$. There will be a phase difference $\Delta\phi \in [0, \pi]$ between these two orthogonal components, leading to the following elliptic equation

$$\frac{\vec{E_x}^2}{E_{x0}} + \frac{\vec{E_y}^2}{E_{y0}} - \frac{2\vec{E_x}\vec{E_y}}{E_{x0}E_{y0}}\cos(\Delta\phi) = \sin^2(\Delta\phi),$$

where $E_{x0}$ and $E_{y0}$ are amplitudes of $\vec{E_x}$ and $\vec{E_y}$. According to the value of $\Delta\phi$, the polarization of $\vec{E}$ can be divided into the following three categories:

**When $\Delta\phi = 0$ or $\pi$**: we have $\vec{E_y} = \pm\frac{E_{y0}}{E_{x0}}\vec{E_x}$, so the signal is linear polarized (LP), as shown in Fig. 4(a). The polarization direction hinges on $\pm\frac{E_{y0}}{E_{x0}}$, the ratio of $\vec{E_x}$ and $\vec{E_y}$.

**When $\Delta\phi = \pm\frac{\pi}{2}$**: we have $\vec{E_x}^2 + \vec{E_y}^2 = \vec{E}^2$, and now the signal is circular polarized (CP), as Fig. 4(b) illustrates. Besides, Fig. 4(d) provides another perspective on how the CP signal is decomposed into two LP signals. Depending on whether $\Delta\phi$ is positive or negative, the rotation direction of the CP signal is in either left-hand circular polarization (LHCP) or right-hand circular polarization (RHCP), which are orthogonal and won't interfere with each other.

**When $\Delta\phi$ is Other Values**: the signal is elliptical polarized (EP), as Fig. 4(c) depicts. Similar to the CP signal, the EP signal also can be divided into left-hand or right-hand.

**Impact of Polarization on the Rx**: The polarization of a signal is accorded with that of its transmitting antenna but may change during propagation. To ensure effective reception, it should match the polarization of the receiving antenna, partially at least. Fig. 5 illustrates how polarization mismatch affects the received signal strength (RSS).

For the LP signal and antenna, RSS decreases as the angle of these two polarization directions increases from $0°$ to $90°$. For the CP signal, the signal can be decomposed into two orthogonal LP signals. Thus, the LP antenna can only receive the component whose polarization direction is parallel to itself but loses half of the signal energy. Similarly, the CP antenna can only receive half of the LP signal's energy. However, when LHCP antenna is used to receive RHCP signals or vice versa, RSS is theoretically zero because these two polarizations are orthogonal. That is the reason why Bifrost can eliminate the ambiguity of two FSDM signals radiated from different LWAs.

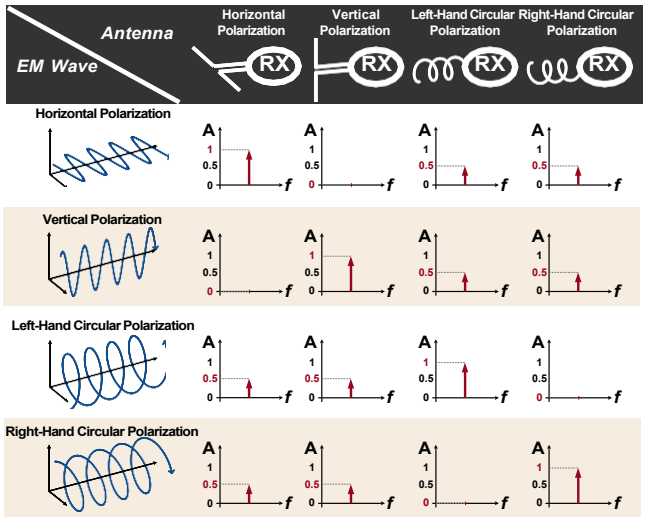

Figure 5: RSS variation according to the polarization of signals and Rx.

## 2.2 Leaky Wave Antenna

LWA belongs to the class of traveling-wave antennas, where the propagating wave inside the antenna structure can "leak" (*i.e.,* radiate) from the waveguide to the free space, hence the name. It can distinctively couple the leaky wave's frequency and radiation direction to produce a frequency and spatial division multiplexing (FSDM) signal, as shown in Fig. 6. Specifically, direction of the signal $\vec{E}_f$ with frequency $f$ can be determined by [71]:

$$\theta(f) = \arccos\frac{\beta(f)}{k_0(f)}, \quad (2)$$

where $\beta(f)$ and $k_0(f)$ are the phase constant along the LWA and the propagation constant in the free space $w.r.t$ $E_f$ [52].

Currently, two main types of LWAs have been extensively studied. 1) *The uniform LWA*, which employs a metallic waveguide with a slit cut along its length [21, 22, 42, 76], as depicted in Fig. 6(b). The FSDM signal leaked from a uniform LWA can only propagate towards the forward region (*i.e.,* [0°, 90°]). 2) *The periodic LWA*, which is typically designed using a dielectric substrate with a periodic array of metal strips (*i.e.,* slots) [10–13] and similar to an antenna array, as shown in Fig. 6(a). The FSDM signal of this type of LWA can propagate towards both forward and backward regions (*i.e.,* [0°, 180°]) [33].

Periodic LWA has been widely studied in recent research due to its versatile slot design and low-cost fabrication using the printed circuit board (PCB) process. These attributes have made it a popular choice in various applications. Bifrost also employs the periodic structure to produce circular polarized signals.

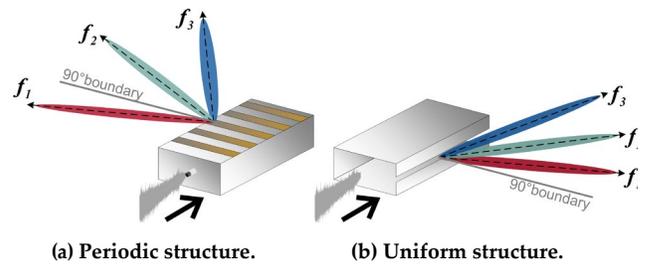

(a) Periodic structure.  (b) Uniform structure.

Figure 6: Typical structures of leaky wave antenna[3].

## 3 BIFROST

In this section, we first articulate how to design the circular polarized LWA (*i.e.,* CPLWA) to transform the input LP signal into the CP signal with the FSDM feature. Then, we present details of our approach of localization with the CPLWA.

### 1.1 CPLWA Design

Unlike many traditional LWAs [10, 13, 21, 22], Bifrost utilizes CP[4] (*i.e.,* RHCP and LHCP) to distinguish different LWAs and corresponding FSDM signals. We specially design a CPLWA that can generate both LHCP and RHCP signals. As shown in Fig. 7(a), our CPLWA has both vertical and horizontal slots to generate orthogonal LP signals, and further to form the CP signal (the bifurcation is designed for performance optimization). According to Eq. (1), a $\frac{\pi}{2}$ phase difference between two LP signals is necessary to generate the CP signal, and this is achieved by adjusting the length of the slots. Denoting the guided wavelength at 5.25GHz of the substrate material is $\lambda_g$, the distance between the center of the horizontal and the vertical slots is $\frac{\lambda_g}{4}$.

In the fabrication process of CPLWA, we adopt a two-layer copper-clad substrate structure, as shown in Fig. 7(b). The substrate material is F4BM-2, whose permittivity $\epsilon$ = 3.02. The top and bottom layers of the substrate consist of copper and have undergone tin immersion plating to prevent oxidation. The bottom layer of copper functions as the ground, and the shorting vias are incorporated to penetrate the substrate, connecting the top and bottom layers in order to ground the top layer. These shorting vias are periodically arranged on the upper and lower boundaries of the substrate and the patch.

The final structure of our proposed CPLWA is depicted in Fig. 7(c), where multiple units are linearly arranged together to enhance the directivity of the FSDM signal, which is similar to the antenna array. Note that a CPLWA is composed of 6 units as an illustration, but 11 units are arranged in practice. This CPLWA features two ports on both ends: one is the feed port that connects to an LP antenna for absorbing the WiFi signal, and the other should connect to a matched 50Ω load. By changing the signal feed port, polarization of the FSDM signal can switch between LHCP and RHCP. If the input signal has gone through all slots and reached the other end, yet still has energy remaining, the matched load will absorb the excess signal.

---
[3]It is worth noting that the 2D radiation pattern is used here for illustration purposes. In reality, the radiation pattern of the leaky wave with a specific frequency is more like a cone, with a generatrix along the propagation direction of the traveling wave.
[4]Unless otherwise specified, CP signals stand for both RHCP and LHCP signals.

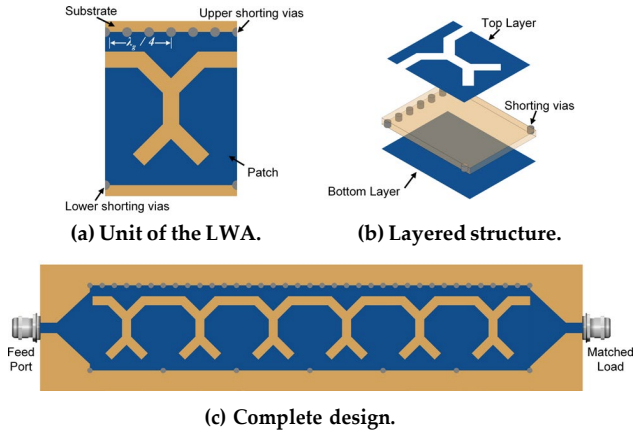

(a) Unit of the LWA.　　(b) Layered structure.

(c) Complete design.

Figure 7: General view of CPLWA used in Bifrost.

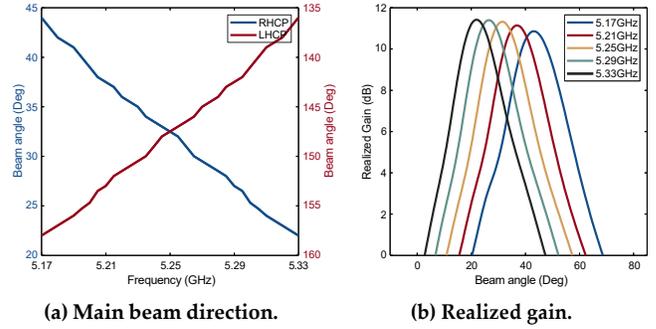

(a) Main beam direction.　　(b) Realized gain.

Figure 8: Key results of the CPLWA.

The CPLWA used in Bifrost is specially designed at 5.17GHz-5.33GHz WiFi band, while this structure and design methodology are universally applicable for other frequencies and bandwidths by properly modifying the relevant parameters.

Now we conduct a quick validation to show the key performance of the proposed CPLWA using ANSYS HFSS. Firstly, the direction of the FSDM signal $w.r.t$ different frequencies is depicted in Fig. 8(a). There is a total 22° field of view (FoV) across the operating frequency band (5.17GHz-5.33GHz). Note that when the LP signal is fed into the right port or left port, the RHCP or LHCP signal will be radiated from 22° to 44° or 136° to 158°, respectively. Fig. 8(b) shows the energy distribution of signals at five different frequencies. It is evident that the energy of the leaky signal concentrates on the correct direction, and their realized gains are all above 11.5dB. Therefore, the direction can be easily identified by examining the energy distribution of signals.

With the proposed CPLWA, we will proceed with elaborating on the core localization algorithm in Bifrost.

### 1.2 Basic Localization Model

Let $S_l$ and $S_r$ respectively denote LHCP and RHCP signals that propagate from corresponding LWAs to the target via the LoS paths. The frequencies of these two signals, $f_l$ and $f_r$, are what we desire for calculating the location. Recall that $S_l$ and $S_r$ are featured in frequency and space division multiplexing (FSDM) and orthogonal CP[5], so these two signals won't interfere with each other. As a result, the target can estimate its relative direction to both LWAs based on the received spectrum and the radiation pattern of the two LWAs. Further, given locations of two LWAs, $L_r (x_r, y_r, z_r)$ of the RHCP LWA and $L_l (x_l, y_l, z_l)$ of the LHCP LWA, the target can output its absolute location. In detail, as we mentioned in §2, the radiation pattern of the LWA is a conical surface at a specific frequency. Therefore, the location $L_t (x_t, y_t)$ of the target device is the intersection point of the two conical surfaces and the horizontal plane of its height. By combining these conditions, $L_t$ can be estimated by solving the following equation set:

$$L_t = (x_t, y_t) = \begin{array}{l} F(L_r, f_r), \\ F(L_l, f_l) \end{array} \quad (3)$$

where $z_t$ is the target's height; functions $F(L_r, f_r)$ and $F(L_l, f_l)$ are mathematical equations of conical surfaces with the location of LWAs as the vertex. These two equations indicate the propagation directions of RHCP and LHCP signals at frequencies $f_r$ and $f_l$, respectively. Taking the RHCP signal as an example, $F = F(L_r, f_r)$ can be formulated as

$$F = (x - x_r)^2 - \frac{(y - y_r)^2}{a^2} - \frac{(z - z_r)^2}{a^2}, \quad (4)$$

where $a = \cot[\theta(f_r)]$.

However, there are two other types of signals impacting the localization accuracy when Bifrost functions: 1) *LP WiFi signal* that is emitted by the WiFi AP, and then received by the target. This signal establishes data communication between the target and the AP and propagates in both the LoS path and multipath. It is also the input signal of LWAs, which will be transformed into FSDM signals by the LWAs. 2) *CP multipath signal* that propagates from LWAs to the target after reflection, resulting in undesired noisy signals at the target.

Thus, we should first identify the frequency of the FSDM signal from the *LP WiFi signal* (discussed in §3.3) and then filter out the *CP multipath signal* as much as possible (discussed in §3.4 and §3.5), to accurately estimate frequencies, $f_l$ and $f_r$, and the target's location.

### 1.3 Identifying Frequencies of CP signals

When Bifrost functions, LWAs need the LP WiFi signal as input, and the target device may also need it for data communication with the WiFi AP. Nevertheless, the LP signal may interfere with the reception of the CP signal, because CP antennas at the target device can receive the LP signal (as already explained in §2). To cancel this interference, we control LWAs to be periodically turned on and off, working in a duty-cycled manner. This design allows the target to identify frequencies that correspond to the CP signal by analyzing the variation in its received spectrum, and at the same time, saves energy of LWAs. Specifically, we exploit WiFi CSI [27, 28, 75, 78] to explore fine-grained information on the amplitude and phase of the subcarriers. Fig. 9 illustrates the result of a proof-of-concept experiment, where subcarriers correspond to LoS and multipath signals are distinguishable in the normalized amplitude of CSI. However, the variation in phase is not obvious, making it challenging to discern useful subcarriers because they are often obscured by random errors and noise. According to this result, we can only extract frequencies of the CP signal based on the amplitude variation in CSI.

---
[5]Unless stated otherwise, CP signals have the property of FSDM.

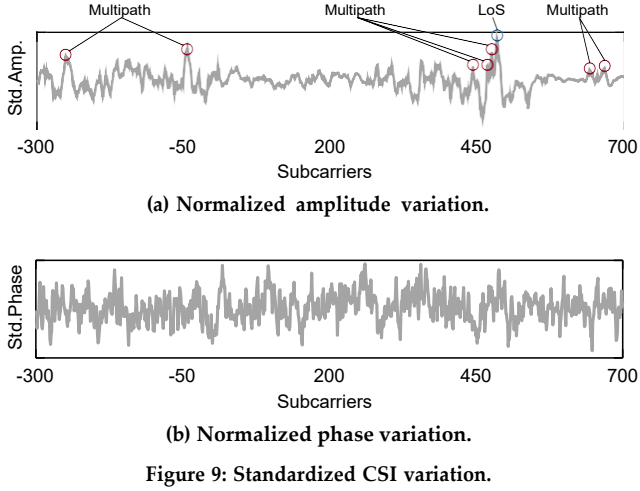

(a) Normalized amplitude variation.

(b) Normalized phase variation.

Figure 9: Standardized CSI variation.

As a LWA turns on or off, we denote the corresponding CSI as $H_{on}(f_k)$ and $H_{off}(f_k)$ for the $k$-th subcarrier with center frequency $f_k$, respectively. The former is jointly influenced by CP and LP signals, while the latter is determined by the LP signal only, leading to the following relationship:

$$\|H_{on}(f_k)\| = \|H^{CP}(f_k) + H^{LP}(f_k)\|,$$
$$\|H_{off}(f_k)\| = \|H^{LP}(f_k)\|, \quad (5)$$

where $\|H^{CP}(f_k)\|$ is the amplitude of subcarriers corresponding to the CP signal, and $\|H^{LP}(f_k)\|$ is that of the LP signal. Based on these two values, we can quantify the variation of CSI caused by the CP signal:

$$\begin{aligned}\|\Delta H(f_k)\| &= \|H^{CP}(f_k)\| \\ &= \|H^{CP}(f_k) + H^{LP}(f_k)\| - \|H^{LP}(f_k)\| \\ &= \|H_{on}(f_k)\| - \|H_{off}(f_k)\|\end{aligned} \quad (6)$$

In order to accurately analyze this variation and mitigate the effect of occasional outliers and noise, a Z-Score normalization procedure is performed on $\|\Delta H(f_k)\|$. We execute a preliminary screening to quickly filter out the subcarriers that are less likely corresponding to the frequencies of the CP signals. A percentage threshold $\varepsilon \in [0, 1]$ is set to select subcarriers with a larger value of $\|\Delta H(f_k)\|$, indicating that these subcarriers undergo significant changes and are more likely to be affected by the CP signal. The value of $\varepsilon$ is chosen empirically based on the degree of multipath. Fig. 10(a) shows a high-level overview of the selected subcarriers, where LHCP and RHCP signals are highlighted in red and blue,

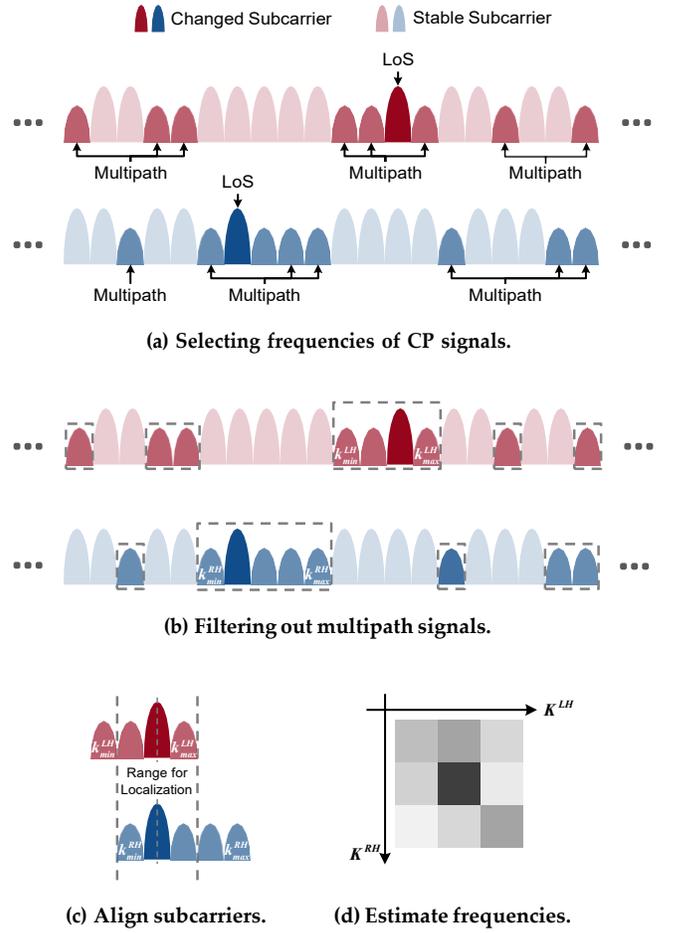

(a) Selecting frequencies of CP signals.

(b) Filtering out multipath signals.

(c) Align subcarriers. (d) Estimate frequencies.

Figure 10: Workflow of selecting correct frequencies (LHCP and RHCP are distinguished by red and blue colors).

respectively. In subsequent stages, we exclusively focus on these selected subcarriers.

### 1.4 Filtering out the Multipath Signal

As shown in Fig. 9(a), even though we have identified the frequencies of the CP signal from the WiFi signal, there still exists the multipath signal, resulting in undesired variation in $\Delta H(f_k)$. Note that the multipath signal is mainly introduced by reflection of the CP FSDM signal. We find that subcarriers corresponding to the multipath signal can be divided into two categories: 1) Sparsely clustered subcarriers $C_s$: FSDM signal with different frequencies and propagation directions may go through reflection at many places, but only a few of those signals reach the target with inconsecutive frequencies, resulting in many sparse clusters of subcarriers[6]. 2) Compactly clustered subcarriers $C_c$: There are some FSDM signals with frequencies close to that of the LoS signal. Those FSDM signals reflect just right near the target device, which will result in a compact and wide cluster of subcarriers influenced by multipath and LoS signals.

---

[6] The polarization of the signals may flip after reflection, and we deal with it as the multipath signal in the frequency domain. Thus, this flip doesn't affect the function of our algorithm.

Here we first try to filter out $C_s$. To do so, all the varied subcarriers are clustered, respectively, as Fig. 10(b) illustrates. Then, the following integral function will be calculated for every cluster to find the one most likely to be corresponding to the LoS signal,

$$C^i = \int_{f^i_{k_{\min}}}^{f^i_{k_{\max}}} \|\Delta H(f^i_k)\| \, df_k \quad (7)$$

where $f^i_{k_{\min}}$ and $f^i_{k_{\max}}$ are the minimum and maximum frequencies of the $i$-th cluster, respectively. The value of $C^i$ can be regarded as the area formed by the curve of $\|\Delta H(f^i_k)\|$ and the two frequencies $f^i_{k_{\min}}, f^i_{k_{\max}}$. The wider the bandwidth and higher the amplitude of a cluster are, the greater the value of its $C^i$ is.

After that, we only retain the cluster that bears the highest $C^i$, which is most likely to be $C_c$ and contains subcarriers corresponding to the LoS signal. However, as we mentioned before, some subcarriers in $C_c$ are also corresponding to the undesired multipath signal. Next, we are going to purify $C_c$ by narrowing down its frequency range as much as possible.

### 1.5 Purifying the LoS Signal for Localization

Denote the frequency range of $C_c$ as $k^r_{\min}, k^r_{\max}$ for RHCP signals and $k^l_{\min}, k^l_{\max}$ for LHCP signals. In both of the two ranges, we are going to find the subcarrier with the largest $\|\Delta H(f_k)\|$ as Fig. 10(c) illustrates. After obtaining them, we denote the index of selected subcarriers as $K^r$ and $K^l$. Next, as Fig. 10(c) depicts, we align $K^r$ and $K^l$, then trim the head and tail to retain the intersection of two clusters, $\|\Delta H^r(f_k)\|$ and $\|\Delta H^l(f_k)\|$. Finally, we multiply $\|\Delta H^r(f_k)\|$ and $\|\Delta H^l(f_k)\|$ to form a weight matrix $G$, which is illustrated in Fig. 10(d).

$$G = \begin{bmatrix} \|\Delta H^r(f_{K^r-\delta})\| \\ \ldots \\ \|\Delta H^r(f_{K^r+\delta})\| \end{bmatrix} \times \begin{bmatrix} \|\Delta H^l(f_{K^l-\delta})\| \ldots \|\Delta H^l(f_{K^l+\delta})\| \end{bmatrix} \quad (8)$$

where $\delta$ is half the length of the vectors $\|\Delta H^r(f_k)\|$ and $\|\Delta H^l(f_k)\|$. Then, we estimate $f_r$ and $f_l$ by computing the weighted average of values in $[f_{K^r-\delta}, f_{K^r+\delta}]$ and $[f_{K^l-\delta}, f_{K^l+\delta}]$, which are weighted by the corresponding values in the matrix $G$. The purpose of this step is still to mitigate the interference of the multipath signal. After that, the estimated values of the two frequencies will be fed into Eq. (4) to output an estimation of the target's location. Note that if there are multiple WiFi links for selection, one can choose the link that results in the smallest size of $\|\Delta H(f_k)\|$, meaning that the range of LoS signals' frequency is reduced to the minimum.

Note that the basis of our localization algorithm is using the different CP signals to distinguish different LWAs, and the CP signals can't be replaced by the LP signals. The reason is that the LP signals may lead to high localization errors or even the breakdown of the localization system. Specifically, once the orientation of LP devices changes, polarization directions of these devices change accordingly. As such, each receiving antenna is very likely to receive FSDM signals from both LWAs and can't distinguish them.

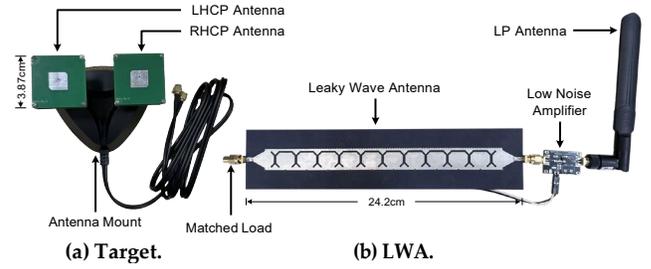

**(a) Target.**   **(b) LWA.**

**Figure 11: Hardware Settings.**

For example, a receiving antenna with 0° polarization can receive both 0° and 90° polarized FSDM signals after rotating 45°. In this case, the target can't distinguish FSDM signals from the two LWAs, and then the localization system can't work. Note that this problem can't be avoided since the target antenna's orientation isn't known in advance. In contrast, CP signals are free from this problem. The RHCP signal can't be received by LHCP antennas no matter which orientation the target antenna has.

Next, we will proceed with describing the prototype implementation to gain insights on the performance of Bifrost in varied settings.

## 4 EVALUATION

We evaluate the performance of Bifrost using two low-cost PCB-based LWAs working at 5.17GHz-5.33GHz and a WiFi sensing platform called PicoScenes [38] to extract CSI. When Bifrost functions, the WiFi transceiver communicates at the same band based on 802.11ax standard [1]. We first describe our implementation and evaluation settings in §4.1. Then, investigation on Bifrost's performance is four-pronged: §4.2 compares Bifrost with SpotFi [43], the state-of-the-art indoor WiFi localization technique, in a real-world indoor setting and NLoS scenarios, and then shows how the localization accuracy can be improved when Bifrost aids SpotFi to function in AP-constrained scenarios; Subsequently, in §4.3, we conduct an ablation study to evaluate the contribution of each sub-module of localization algorithm; Then, in §4.4, we dissect the impacting factors on localization accuracy, including multipath, transmission power, as well as the distance between LWAs and the AP; Also, we evaluate the influence of deploying Bifrost on data communication of WiFi transceivers in §4.5; Finally we summarize the evaluation in §4.6.

### 4.1 Implementation and Experimental Methodology

**Hardware and Software.** Our proposed LWA is shown in Fig. 11(b). The main body of our LWA is 24.2cm× 5.2cm, containing 11 single units designed to ensure most input signals' energy can be leaked out. One of the LWA's feed ports is connected to a LP antenna for receiving the WiFi signal while the other port is connected to a 50Ω matched load to absorb the remaining energy of the signal that goes through the entire LWA structure. By switching the feed port, the polarization of the FSDM signal can be altered between LHCP and RHCP. Besides, a low-noise amplifier powered by a small rechargeable battery is utilized to boost the input signal with 0.43W power consumption. A NE555 timer IC with a load switch circuit is employed to control the on-off state of the amplifier and further LWAs, resulting in a 20% duty-cycled manner for energy saving. The cost of each proposed LWA is 7.41 USD, where 4.36 USD is for the material cost and 3.05 USD is for the control circuit. To receive the CP FSDM signal, we equip the target with two 3.87cm$\times$3.87cm patch antennas, as Fig. 11(a) depicts. One antenna is LHCP, while the other is RHCP, and both are fixed on the antenna mount connected to COMFAST AX210 WiFi card [17] on the host computer.

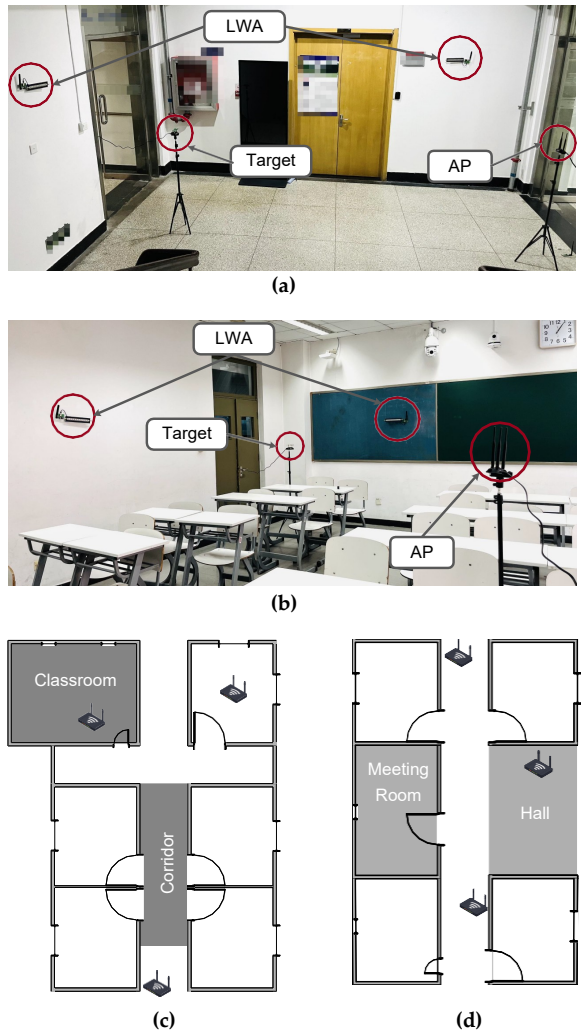

**Figure 12: Experimental scenarios and deployment: (a) The hall scenario; (b) The classroom scenario; (c) The APs' deployment in the corridor and the classroom; (d) The APs' deployment in the hall and the meeting room.**

We use PicoScenes, a WiFi sensing platform, to send WiFi packets at AP with 20dBm, and extract CSI at the target. In the working band of Bifrost, PicoScenes can procure CSI data of 2025 subcarriers with indexes [-1012, 1012][7]. We run PicoScenes on Ubuntu 20.04,

---

[7] PicoScenes automatically interpolates the *0-th* and other *32 pilot subcarriers* besides 1992 tone RUs in this band.

then analyze CSI data and execute the localization algorithm on MATLAB 2022b.

**Baseline.** We compare Bifrost with SpotFi, the state-of-the-art indoor WiFi localization technique, under various settings. To ensure the validity of our results, we make our best effort to re-implement SpotFi and ensure fairness through comparison. We evaluate the performance of SpotFi by deploying multiple WiFi APs *strictly based on the real-world settings of WiFi APs*, as Fig. 12 shows. Before each set of experiments, we use a laser rangefinder to obtain the ground-truth, including coordinates of the target device and LWAs.

**Scenarios and Deployment.** We select four typical indoor scenarios for evaluation, across different sizes and different levels of multipath effect: 1) A small-size hall (6.2m$\times$4.5m) with few multipath; 2) A long and narrow corridor (7.5m$\times$2.1m) with few multipath; 3) A small-size meeting room (5.7m$\times$4.9m) with rich multipath; 4) A large-size classroom (10.6m$\times$7.1m) with rich multipath. In each scenario, two LWAs are attached to two orthogonal walls. The target device is mounted onto tripods, keeping the height constant across all experiments.

### 4.2 Overall Performance

In this section, we first evaluate the localization accuracy of Bifrost and SpotFi in real-world settings, where WiFi APs in experiments are deployed at the same positions as those in practice. Then we deploy Bifrost in the meeting room and classroom, where SpotFi doesn't work well, to enhance the performance of SpotFi, so as to see the accuracy improvement brought by Bifrost.

**Performance Comparison in Realistic Settings.** In reality, most indoor WiFi APs are dispersively deployed at different locations and very likely separated from each other by walls so that LoS paths are usually obstructed. Thus, the target device is hard to establish more than one LoS connection with APs, according to our real-world investigation (*i.e.,* Fig. 2). We evaluate the performance of SpotFi in these practical indoor settings, and also the localization error of Bifrost when deployed in the above-mentioned four scenarios. 50 locations are chosen in each scenario for location estimation. The evaluation results are reported in Fig. 13 (The solid blue line stands for Bifrost and the dashed red line stands for SpotFi).

In the hall, both Bifrost and SpotFi are supposed to exhibit the best performance due to the low-level multipath effect, but the median error of SpotFi is 1.23m, which is more than 2$\times$ of Bifrost's 0.61m. This is because only one decent LoS signal can be obtained at most locations due to the blockage of walls even though three APs are deployed around. As the pie chart illustrates, SpotFi outperforms Bifrost at only 9 locations. When it comes to the corridor scenario, the median error of SpotFi increases to 1.77m because two of the three APs are situated inside rooms so that AoAs obtained by the target are heavily distorted. We note that the median error of Bifrost also increases to 0.76m. This slight performance degradation is mainly due to the extension of the localization range, which is further investigated in §4.4.

Next, we switch to the meeting room where more pronounced multipaths exist. What's worse, there is no AP in the meeting room, more challenging for both two approaches to function. The accuracy of the two approaches is unsurprisingly degraded, where the median error is 1.95m in SpotFi and 0.91m in Bifrost. Similarly, the

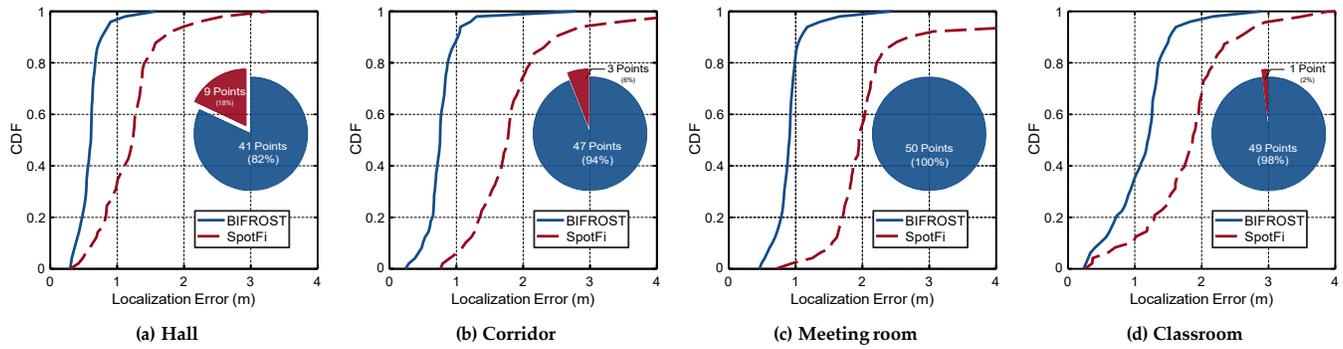

**Figure 13: Overall performance of Bifrost and SpotFi across different scenarios (The pie charts represent how many locations where each method shows a lower error).**

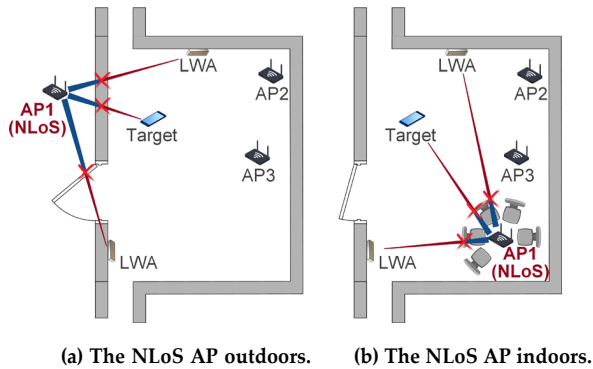

**(a) The NLoS AP outdoors.** **(b) The NLoS AP indoors.**

**Figure 14: Deployment of the NLoS settings.**

performance of SpotFi is restrained due to the lack of the LoS signal. Bifrost exhibits acceptable performance in this tough environment and avoids escalation of errors. This can be attributed to two aspects. On one hand, Bifrost can function once the input signal has enough energy, without the need for LoS AP. On the other hand, Bifrost exploits a delicate algorithm to tame the multipath effect. We will further discuss issues of multipath and NLoS in §4.4. In this scenario, SpotFi doesn't outperform Bifrost on any point.

Finally, we set SpotFi and Bifrost in the large-size classroom with rich multipath. With a LoS AP, the median error of SpotFi is reduced to 1.87m, which is better than that in the meeting room with no LoS AP. By contrast, the median error of Bifrost increases to 1.20m, mainly due to a longer distance between LWAs and WiFi APs and more multipath.

Through all experiments in four scenarios, the median error of Bifrost is 0.81m, which is 52.35% less than that of SpotFi (*i.e.*, 1.70m). Bifrost outperforms SpotFi at most locations, except at which the target can obtain 3 LoS signals from 3 APs. However, as shown in Fig. 13, the chance for SpotFi to achieve better performance is less than 7%.

**Performance Comparison in NLoS Scenarios.** Then we conduct two groups of experiments to demonstrate Bifrost's ability of localization in NLoS scenarios and compare its performance with that of SpotFi.

In the first group of experiments, we deploy the localized target and the LWAs in a hall. As Bifrost only uses one AP to function, we evaluate the performance of Bifrost when this AP is inside and outside the hall (*i.e.*, LoS and NLoS scenarios). The results in Fig. 15 show that the median errors of Bifrost are 0.61m in LoS and 0.73m in NLoS, respectively. Meanwhile, in the same hall, we also evaluate the performance of SpotFi in LoS and NLoS scenarios, respectively. In the LoS scenario, 3 APs are deployed in the hall and can establish LoS connections with the target. In the NLoS scenario, as Fig. 14(a) shows, one of the APs (*i.e.*, AP1) is outside the room, while the other 2 APs (*i.e.*, AP2 and AP3) can connect with the target along the LoS paths. We find that the median error of SpotFi increases from 0.45m in LoS to 1.15m in NLoS. The error may further go beyond 1.6m if only one AP is left in LoS, as reported in [43].

In the second group of experiments, we compare the performance of Bifrost and SpotFi using a different NLoS setting. As Fig. 14(b) shows, we deploy the localized target, LWAs and three APs in the same hall. One of the three WiFi APs (*i.e.*, AP1) is deliberately deployed around the corner and surrounded by multiple chairs, so it can't establish LoS connections between the target or the LWAs, while the other 2 APs (*i.e.*, AP2 and AP3) can connect with the target along the LoS paths. SpotFi uses all 3 APs to localize the target, and its median error is 1.21m. Bifrost only uses the AP in NLoS (*i.e.*, AP1) to function, and its median error is 0.69m, which is 42.98% less than that of SpotFi.

These two groups of experiments demonstrate that Bifrost provides relatively stable performance when the WiFi AP is in LoS and NLoS scenarios. In NLoS scenarios, Bifrost can achieve much more accurate performance than SpotFi.

**Performance Enhancement when Bifrost Aids SpotFi.** Next, we deploy Bifrost where SpotFi shows poor accuracy to see if Bifrost can aid SpotFi to improve localization accuracy. Actually, it is impossible to deploy Bifrost everywhere, so we choose the meeting room and classroom where localization accuracy is heavily affected by constrained APs and reports the worst results. Specifically, when the target gets into these two scenarios, its location will be reported by Bifrost. Otherwise, the target keeps using SpotFi for indoor localization.

As shown in Fig. 16, the median localization error is 1.13m when Bifrost aids SpotFi, achieving 33.54% error reduction compared with SpotFi operating independently in all scenarios. This indicates that Bifrost can not only work independently, but also enhance localization accuracy of existing localization techniques.

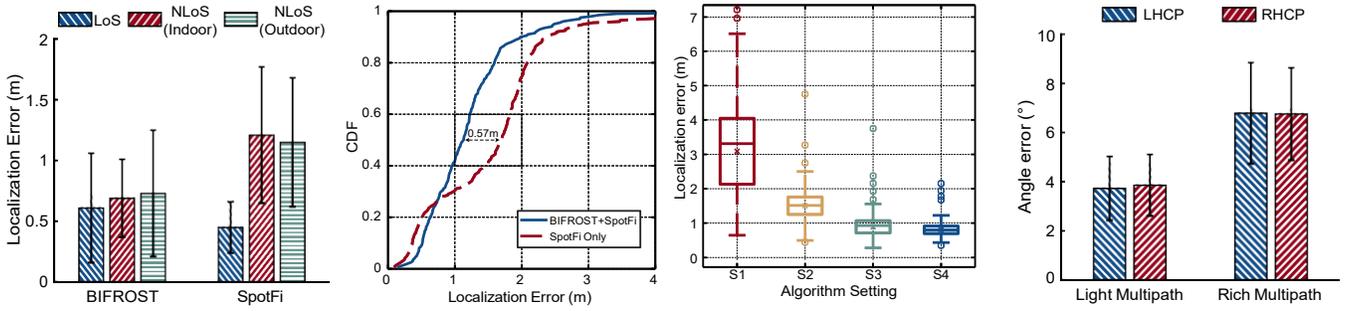

Figure 15: Performance of Bifrost and SpotFi in the NLoS scenario.

Figure 16: Performance enhancement brought by Bifrost.

Figure 17: Ablation study on the localization algorithm.

Figure 18: Impact of the multipath effect.

### 4.3 Ablation Study

There are three crucial sub-modules in Bifrost's localization algorithm, that is, identifying the frequencies of CP signals (*module 1* presented in §3.3), filtering out the multipath signal (*module 2* presented in §3.4) and purifying the LoS signal for localization (*module 3* presented in §3.5). We conduct an ablation study to evaluate the contribution of each sub-module to localization accuracy. The evaluation is conducted under four settings, S1: without any sub-module, S2: only with module 1, S3: with modules 1 and 2, and S4: with all three modules.

Fig. 17 reports the results of this ablation study. If we do nothing and directly extract frequencies from raw amplitude data of CSI, the median localization error will surge to 3.31m (S1). Instead, once the LP WiFi signal is filtered out, the frequencies of CP signals can be highlighted, which results in the median localization error of 1.51m (S2). Further, the results of S3 and S4 show that the median error will be reduced to around 0.93m and 0.81m if we filter out the multipath signal and purify the LoS signal. These results show the necessity and contribution of each module in our design.

### 4.4 Impacting Factors

Next, we analyze the impact of three different factors on the performance of Bifrost, that is, multipath in the environment, the transmission power, as well as the distance between LWAs and WiFi AP.

**Multipath.** We examine the AoA estimation accuracy of Bifrost in multipath scenarios. We fix the positions of LWAs and the target, then change the number of indoor objects (*i.e.,* chairs and desks) to create different degrees of multipath. Specifically, two desks are first set in the room to emulate a light multipath environment, and then ten chairs are further added to produce richer signal reflections. The results in Fig. 18 indicate that the AoA estimation accuracy degrades as the multipath is intensified, where the median angle error initially sits around 3.8°, and then increases to around 6.7°. The more multipath exists, the more sparsely clustered subcarriers $C_s$ are formed. Thus, when these clusters are stacked with each other to form a wider cluster, there is a certain chance for our algorithm to misidentify the wrong LoS signal, causing greater errors in AoA estimation.

We also note that Bifrost maintains relatively stable performance across different polarizations. The difference between median errors of LHCP and RHCP signals is less than 0.3°, which underscores the robustness of our proposed LWA and localization algorithm.

**Transmission Power.** The default transmission power of AP is 20dBm in our above-mentioned evaluations, and we now vary this value to investigate its impact on localization performance. Moreover, as mentioned before, we can't always guarantee that the WiFi AP establishes LoS path with LWAs, so we also compare the situation of the AP at LoS and NLoS scenarios in each setting of transmission power. We place AP at 2m distance outside the door and the target 2m inside the door, switching between the LoS and NLoS scenarios by opening and closing the door. Results in Fig. 19 show that decreasing the transmission power leads to an increase in the localization error, regardless of whether the AP is at LoS or NLoS. Besides, the errors in LoS scenario are always lower than that of NLoS for the same transmission power. These findings indicate the negative impact on localization performance that NLoS can have.

However, we also observe that as the transmission power increases, the impact of NLoS on the performance of Bifrost decreases, albeit gradually. Notably, when the transmission power is set at 20dBm, the median errors are 0.61m and 0.73m at LoS and NLoS scenarios, respectively. In practical scenarios, this performance is sufficient to meet the requirements of most location-based applications.

**Distance between AP and LWAs.** The performance of Bifrost may be influenced by the energy of the input WiFi signal fed into LWAs, because it determines the SNR (signal-to-noise ratio) of the FSDM signal. The energy of the input WiFi signal is mainly related to two factors, namely the transmission power and the distance between the AP and LWAs. While the former factor is previously discussed, we here probe into the impact of distance. We carry out the experiments along the corridor and remove the reflectors as far as possible, while the distance is set to 2.5m, 5m, 7.5m, and 10m. Results in Fig. 20 demonstrate that the localization error increases with distance and may even result in outliers. The median errors are 0.63m, 0.65m, and 0.93m in the first three groups of experiments, all of which are below 1m, yet spike to 1.49m in the setting of 10m distance. Despite this, the range of 7.5m is sufficient to cover most rooms in a typical building, thus ensuring the feasibility of Bifrost's function.

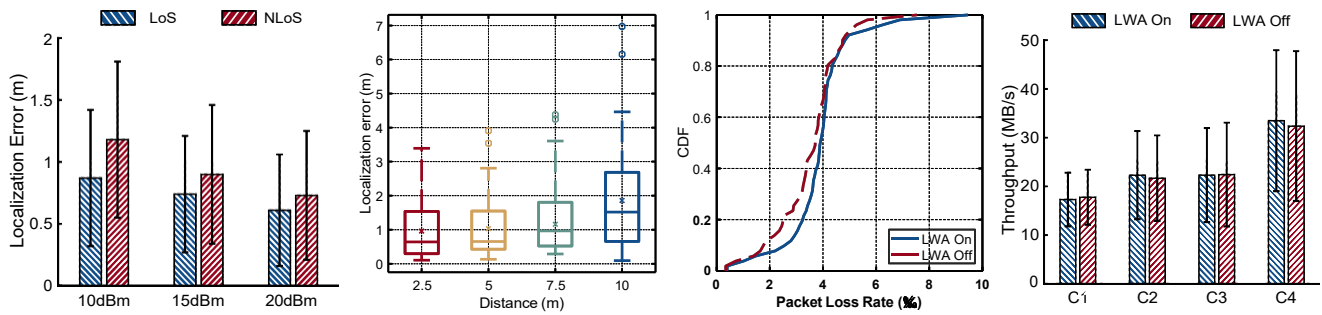

Figure 19: Impact of the transmission power.
Figure 20: Impact of the distance between AP and LWAs.
Figure 21: Impact on the AP and the target of Bifrost.
Figure 22: Impact on other WiFi connections.

### 4.5 Impact on Communication

In this section, we evaluate the impact of deploying Bifrost on the WiFi connections, including the connection between the AP and the target as well as other connections. Firstly, we control the AP to transmit 1000 packets at a 50 *ms* interval, and the packet loss rate is recorded in each group of experiments. The results in Fig. 21 show that the median packet loss rates are 3.92‰ and 3.71‰ when the LWA is on and off, respectively. This 0.2 ‰ difference implies that the function of Bifrost has a negligible influence on the AP-target communication.

Secondly, we place Bifrost's transceiver at an intersection region covered by two commercial APs (AP1 in a classroom and AP2 in a laboratory) with good signal quality. We then use different off-the-shelf smartphones to establish WiFi connections with these APs and record the variation in throughput over 2 hours for each connection (C1: OnePlus 9-AP1, C2: iPhone 13-AP2, C3: OnePlus 9-AP1, and C4: iPhone-13-AP2). The results are shown in Fig. 22. We find that the median throughput degrades 2.7% and 0.4% in C1 and C3, which have nearly no impact on the network quality or user experience. Interestingly, the throughput increases when the LWAs are turned on for C2 and C4. We attribute this increase to the statistical error that is mainly caused by changes in network quality and wireless channels.

### 4.6 Summary of Evaluation

Based on the above evaluations on Bifrost, the following summary can be drawn:

1) The median localization error of Bifrost is 0.81m, which is 52.35% less than that of SpotFi in arguably realistic indoor settings.
2) Bifrost can be deployed in scenarios without enough APs to help SpotFi enhance performance, reducing the overall localization error of SpotFi by 33.54%.
3) Distance between LWAs and APs, multipath and transmission power influence Bifrost's performance differently, yet the absolute accuracy never degrades drastically.
4) The deployment of Bifrost has a negligible impact on the communication quality of either the link between the AP and the target or other WiFi connections.

## 5 DISCUSSION

In this section, we discuss practical issues concerning the applicability and efficacy of Bifrost.

**Complexity of Deployment.** Deploying Bifrost can be easy and straightforward via two steps: stick LWAs to the wall, and measure LWAs' coordinates. Compared with most existing indoor localization methods, Bifrost works in a plug-and-play manner, requiring neither complex configurations nor additional operations on APs and the target.

**FoV and Coverage of LWAs.** Bifrost achieves 22° FoV in the current prototype by using 160MHz bandwidth (5.17GHz - 5.33GHz). The FoV and coverage can be expanded by using the entire WiFi band, including frequencies at 2.4GHz, 5.2GHz, and 5.8GHz [47]. This expansion is feasible because most existing WiFi devices have supported dual- or tri-band functionality.

**Applicability.** Considering that most of the current commercial WiFi devices are equipped with LP antennas, they may be not compatible with Bifrost yet. There are two potential solutions to enhance the applicability of Bifrost. On one hand, some commercial off-the-shelf CP antennas (*e.g.,* CP flat patch antennas [45] of L-com, Inc) are developed to be integrated with existing WiFi APs. Bifrost can be deployed on such devices. On the other hand, in our future work, we will study how to utilize LP rather than CP signals to improve the applicability of Bifrost. To distinguish LWAs using the LP signals, different phase shifts or OOK patterns may be exploited.

Besides, the indoor obstacles may also influence the applicability of Bifrost. The reason is that the localization performance will degrade if the LoS paths between LWAs and the target are blocked by the obstacles. Therefore, one may select proper positions to deploy LWAs to avoid NLoS propagation to the target to be localized. However, the LoS path between LWAs and the WiFi AP isn't a precondition. As long as the LWAs can receive the signal from the WiFi AP, Bifrost can work.

**Lifetime and Maintenance Cost.** The rated current of LWAs is 0.86mA. A LWA is powered with a 1600mAh battery and works at 20% duty cycle. So the estimated lifetime of a LWA is over 9302 hours (≈387 days) and the maintenance cost is recharging the battery once every 387 days.

**Potential Interference.** One may be concerned that if multiple LWAs are deployed closely, LWAs with the same polarization will interfere with each other. However, each room only has one RHCP LWA and one LHCP LWA in the setting of Bifrost, so LWAs with the same polarization are separated by walls. Interference signals must propagate through the wall, after which they only have low strength. Therefore, different pairs of LWAs hardly interfere with each other.

## 6 RELATED WORK

In this section, we briefly summarize existing works in the fields related to our work.

### 6.1 Application of LWA

The work closest to ours is 123-LOC [42], which presents a THz LWA with two perpendicular slots to radiate horizontal and vertical polarized FSDM signals. Range and angle estimation is then performed by the receiver based on the bandwidth and frequencies of received signals. In comparison, Bifrost reduces the impact of multipath and achieves room-scale localization, which is a challenging task for THz signals.

LeakyTrack [21] tracks the object between two LWAs based on the observation that nodal and environmental motion changes the received spectral profile of FSDM signals. [76] investigates the security of THz networks with LWAs and shows that FSDM signals of the LWA can hinder eavesdroppers, *e.g.,* by using a wide-band transmission. [20] and [22] study single-shot link discovery with the help of FSDM signals from the LWA. A receiver can discover the direction of the path from the transmitter in one shot. In contrast to those works that require a specific feeding device for THz LWA, Bifrost operates in the WiFi band and works in a plug-and-play manner, providing better applicability and convenience. Additionally, Bifrost addresses relevant challenges, including multipath, noise and ambiguity, by delicately designing the hardware and localization algorithm.

### 6.2 WiFi-based Indoor Localization

There have been numerous efforts on indoor localization with WiFi [16, 49, 61, 68–70, 84]. Traditional fingerprint-based techniques have been widely used by mapping the RSS readings from multiple APs with locations [46, 66]. Techniques based on AoA and ToF have become more prevalent recently. For example, ArrayTrack [69] proposes an AoA-based WiFi localization system that incorporates multiple APs and the Multiple Signal Classification (MUSIC) algorithm. SpotFi [43] proposes a MUSIC algorithm to obtain AoA and ToF simultaneously. The $M^3$ system [16] reduces the amount of APs to only one by utilizing multipath signals and frequency hopping among multiple channels.

Despite such inspiring advances, the existing proposals may chop up the communication link between the target and the AP when the target hops between different APs or channels. In contrast, Bifrost does not interfere with the communication link, which supplements the APs' localization ability, without compromising their communication ability.

### 6.3 Polarization of the Wireless Signal

LLAMA [15] designs a metasurface to mitigate polarization mismatch by rotating the polarization of wireless signals, which is achieved by applying the bias voltage to the orthogonal components (like $\vec{E_x}$ and $\vec{E_y}$ shown in Fig. 4) of input signals. RoS [55] and mmTag [51] propose well-designed Van Atta arrays. They all change the polarization of input mmWave signals to the orthogonal one to deal with the self-interference between the incoming signals and the backscattered signals. IntuWition [77] observes that different materials can reflect and scatter the incoming polarized signals in different ways, based on which it exploits the technique to classify various materials. SiWa [83] utilizes the similar principle to inspect the wall structure without undermining the structural integrity.

The above-mentioned works mainly focus on mutable LP signals. Bifrost instead explores the use of orthogonal CP signals, providing more robust performance.

### 6.4 Backscatter-aided Localization

Enabled by the backscatter technology [8, 26, 29–31, 37, 53, 55, 58], many novel applications are enabled, one of which is localization. Both Hawkeye [8] and Millimetro [58] design backscatter tags based on Van Atta arrays to enhance the energy of backscatter signals, so they can localize tags in long range (over 100m). By assigning unique OOK modulation frequencies to different tags, those two works can also identify and localize tags simultaneously. Moreover, RFID technology [34–36, 39–41] has been widely used in localization tasks. As a typical backscatter technology, RFID can modulate information via the RFID tags. Then, RFID reader can usually infer the range or orientation to the tags by analyzing the phase variation of the backscatter signals.

Compared to those works, Bifrost utilizes tags (*i.e.,* LWAs) to create FSDM signals to localize another target, rather than the tag itself.

## 7 CONCLUSION

This paper introduces Bifrost, a low-cost and plug-and-play technique to enhance the availability and accuracy of WiFi localization. It can either aid existing techniques to improve their performance, or operate independently to outperform the state of the arts in arguably realistic indoor settings, without affecting ongoing data communication of WiFi networks. What sets Bifrost apart from other solutions is the exploration in the polarization of wireless signals and the dispersion property of LWAs, which embodies the concept of RF computing [15, 29, 53, 55]. We plan to explore the research space further in this direction.

## ACKNOWLEDGMENTS

We thank our anonymous shepherd and reviewers for their insightful comments. This work is partially supported by the National Natural Science Foundation of China under grant No. U21B2007, and the Guoqiang Institute of Tsinghua University under grant No. 2021GQG1002.